\documentclass[%
 aps,prb,
 amsmath,amssymb,
 preprint,%
]{revtex4-2}
\usepackage[pdftex]{graphicx}
\usepackage{dcolumn}
\usepackage{bm}
\usepackage[T1]{fontenc}
\usepackage{mathptmx}

\begin{document}
\title[]{Dipole Scattering at the Interface: The Origin of Low Mobility observed in SiC MOSFETs}


\author{Tetsuo Hatakeyama}
\affiliation{Toyama Prefectural University,  5180 Kurokawa, Imizu, Toyama 939-0398, Japan}
\author{Hirohisa Hirai}%
\affiliation{Advanced Power Electronics Research Center, 
National Institute of Advanced Industrial Science and Technology (AIST), 16-1 Onogawa, Tsukuba, Ibaraki 305-8569, Japan}

\author{Mitsuru Sometani}
\affiliation{Advanced Power Electronics Research Center, 
National Institute of Advanced Industrial Science and Technology (AIST), 16-1 Onogawa, Tsukuba, Ibaraki 305-8569, Japan}

\author{Dai Okamoto}
\affiliation{Toyama Prefectural University,  5180 Kurokawa, Imizu, Toyama 939-0398, Japan}

\author{Mitsuo Okamoto}
\affiliation{Advanced Power Electronics Research Center, 
National Institute of Advanced Industrial Science and Technology (AIST), 16-1 Onogawa, Tsukuba, Ibaraki 305-8569, Japan}

\author{Shinsuke Harada}
\affiliation{Advanced Power Electronics Research Center, 
National Institute of Advanced Industrial Science and Technology (AIST), 16-1 Onogawa, Tsukuba, Ibaraki 305-8569, Japan}

\date{\today}

\begin{abstract}
In this work, the origin of the low free electron mobility in SiC MOSFETs is investigated
using the scattering theory of two-dimensional electron gases.
We first establish that neither phonon scattering nor Coulomb scattering can be the 
cause of the low observed mobility in SiC MOSFETs;
we establish this fact by comparing the theoretically calculated mobility considering
these effects with experimental observations.
By considering the threshold voltages and the effective field dependence of 
the mobility in SiC MOSFETs,
it is concluded that the scattering centers of the dominant mechanism are electrically neutral and 
exhibit a short-range scattering potential.
By considering charge distribution around a neutral defect at the interface, it is established that 
an electric dipole induced by the defect can act as a short-range scattering potential.
We then calculate the mobility in SiC MOSFETs assuming that there exist a high density of dipoles at the interface.
The calculated dipole-scattering-limited mobility shows 
a similar dependence on the effective field dependence to that observed in experimental results.
Thus, we conclude that scattering induced by a high density of electric dipoles at the interface 
is dominant cause of the low mobility in SiC MOSFETs.
\end{abstract}
\maketitle

\newcommand{\Ec}{E_\mathrm{C}}
\newcommand{\Dit}{D_\mathrm{it}}
\newcommand{\Na}{N_\mathrm{A}}
\newcommand{\nfree}{n_{\mathrm{s}}^\mathrm{free}}
\newcommand{\ntrap}{n_{\mathrm{s}}^\mathrm{trap}}
\newcommand{\ntotal}{n_{\mathrm{s}}^\mathrm{total}}
\newcommand{\rfree}{R^\mathrm{free}}
\newcommand{\mufe}{\mu_\mathrm{FE}}
\newcommand{\mueff}{\mu_\mathrm{eff}}
\newcommand{\Emax}{E_\mathrm{max}}
\newcommand{\Vmax}{V_\mathrm{max}}
\newcommand{\re}{\mathrm{e}}
\newcommand{\br}{\bm{r}}
\newcommand{\bk}{\bm{k}}
\newcommand{\ptsc}{\phi_\mathrm{scC}}
\newcommand{\ptSC}{\bar{\phi}_\mathrm{scC}}
\newcommand{\ptC}{\phi_\mathrm{C}}
\newcommand{\ptscC}{\phi_\mathrm{scC}}
\newcommand{\kb}{k_\mathrm{B}}
\newcommand{\Ns}{n_\mathrm{s}}
\newcommand{\Eeff}{E_\mathrm{eff}}
\newcommand{\epss}{\varepsilon_\mathrm{SiC}}
\newcommand{\muH}{\mu_\mathrm{H}}
\newcommand{\muF}{\mu_\mathrm{free}}
\newcommand{\muph}{\mu_\mathrm{ph}}
\newcommand{\muC}{\mu_\mathrm{C}}
\newcommand{\mudp}{\mu_\mathrm{dp}}
\newcommand{\mudpph}{\mu_\mathrm{phdp}}
\newcommand{\muphdp}{\mu_\mathrm{phdp}}
\newcommand{\muSR}{\mu_\mathrm{SR}}
\newcommand{\Ndpl}{n_\mathrm{dpl}}
\newcommand{\Vth}{V_\mathrm{th}}
\newcommand{\rtd}{r_\mathrm{c}}
%
%
\newcommand{\brc}{\bm{r}_\mathrm{c}}
\newcommand{\rc}{r_\mathrm{c}}
\newcommand{\bpdp}{\bm{p}_\mathrm{dp}}
\newcommand{\Ndp}{N_\mathrm{dp}}
\newcommand{\pdp}{p_{\mathrm{dp}}}

\newpage

In recent years, the adoption of ultralow-loss silicon carbide (SiC) metal-oxide-semiconductor field-effect transistors (MOSFETs)
has become common in advanced power electronic systems, 
such as in the traction system of new types of trains and electric vehicles \cite{jrtokai,EVzhu}.
However, even in the case of modern SiC MOSFETs, the low mobility in the MOS channel 
significantly affects their performance \cite{Kimoto_2020}. 
A technological breakthrough that enhances the mobility in the MOS channel is of great importance to further improvements in the performance of MOSFETs. To address this issue, many studies have been devoted to 
the causes of the low observed mobility in SiC MOSFETs \cite{cooper,afanasiev_pss,saks2000,deak_iop2007}.
It has been established that there are two causes of the low mobility in SiC MOSFETs \cite{saks2000,Hatakeyama_2017,Hatakeyama_2019,Masuda_2019,Hirai_2019,Sometani_2019,Hirai_2020}.
One cause is the reduction in the free electron density ($\nfree$) in the MOS channel 
caused by a high density of interface states ($\Dit$) near the bottom of  
the conduction band ($\Ec$) \cite{saks2000,Hatakeyama_2017,Hatakeyama_2019, Masuda_2019}. 
The second cause is the low free electron mobility ($\muF$) of the remaining mobile electrons in the MOS
channel \cite{Hatakeyama_2017,Hatakeyama_2019,Hirai_2019,Sometani_2019,Hirai_2020,Noguchi_2017,Noguchi_2019,Noguchi_2020}. 
It has been found that the high density of $\Dit$ near $\Ec$ can be reduced and thus the $\nfree$ can be improved \cite{Hatakeyama_2017,Hatakeyama_2019,Masuda_2019}.
However, the possible increase of the $\muF$ is limited \cite{Hatakeyama_2017,Hatakeyama_2019,Masuda_2019}.
Even in the case of SiC MOSFETs on 4H-SiC$(0 \bar{3} 3\bar{8})$ faces annealed in nitric oxide, 
the $\muF$ is only \textasciitilde 100 cm$^2$/Vs \cite{Hatakeyama_2019,Masuda_2019},
which is much smaller than the mobility expected based on the bulk mobility (\textasciitilde 1000 cm$^2$/Vs) \cite{Cree_1994,Hatakeyama_2003,Matsuura_2004}. 
For further improvements in the performance of SiC MOSFETs,
it is essential to identify the origin of the low  $\muF$.
\par
Owing to its technological importance, a number of detailed experimental studies 
on the $\muF$ have been undertaken using Hall effect measurements \cite{Hirai_2019,Sometani_2019,Hirai_2020,Noguchi_2017,Noguchi_2019,Noguchi_2020}.
 We note that the experimental results regarding the mobility in MOSFETs
 are generally parameterized bt the Effective field ($\Eeff$) in order to investigate
 the mobility-limiting scattering mechanisms \cite{Stern_1967,Ando_1982,Takagi_1994_1,Takagi_1994_2}.  
 The $\Eeff$ is defined as follows:
 \begin{equation}
 \Eeff=\frac{e}{\epss}\left(\Ndpl+\eta \nfree \right),
 \label{eq:Eeff}
 \end{equation}
 where $\Ndpl$, $\epss$, and $\eta$ are the area density of ionized acceptors in
 the depletion region, the permittivity of SiC, and an empirical dimensionless constant, respectively \cite{Takagi_1994_1}.
 When considering electrons, $\eta$ is assumed to be 1/3 \cite{Stern_1967,Ando_1982,Takagi_1994_1}. 
 To simplify the discussion,the Hall mobility ($\muH$) is regarded to be equal to the $\muF$ \cite{Hirai_2019}.
 In 2017, Noguchi {\it et al.} reported on the dependence of the $\muH$ in SiC MOSFETs 
 on the $\Eeff$ \cite{Noguchi_2017}.
 They found that the $\muF$ of a SiC MOSFET on low-doped ($\Na=3\times 10^{14}$ cm$^{-3}$) P-well 
 is roughly proportional to $\Eeff^{-0.39}$, where $\Na$ denotes the acceptor density of the P-well .
 They regarded this $\muF$ as being phonon-scattering-limited mobility ($\muph$)
 because the obtained $\Eeff$ exponent was close to that expected for $\muph$,
 which is $-$1/3 \cite{Takagi_1994_1,Price_1981}.
 However, the maximum value of their reported mobility was less than 1/3 of the bulk mobility of 4H-SiC.
 By contrast, Sometani {\it et al.} reported that the $\muF$ in SiC MOSFETs on an ultra-low-doped ($\Na < 1 \times 10^{14}$) 
 P-well was comparable to the bulk mobility of 4H-SiC \cite{Sometani_2019}.
 They also reported that the $\muF$ exhibited the same temperature dependence as the bulk mobility, 
 which is limited by  phonon scattering, over a wide range of temperatures.
 They hypothesized that the decrease in the $\muF$ on a low-doped P-well 
 compared with that observed on an ultralow-doped P-well is caused by the scattering centers located
 at the SiO$_2$/SiC interface.
 They referred to these unidentified scattering centers as unidentified scattering origin (USO).
 In this study, we verify their finding via theoretical calculations of $\muF$
 based on the scattering theory of quasi-two-dimensional electron gases (2DEGs) \cite{Ando_1982,Masaki_1989}.
 Furthermore, we identify the dominant scattering mechanism, or USO, by
 comparing experimental results and theoretical calculations.\par
\begin{figure}[htbp]
\centering
\includegraphics[width=8.6cm]{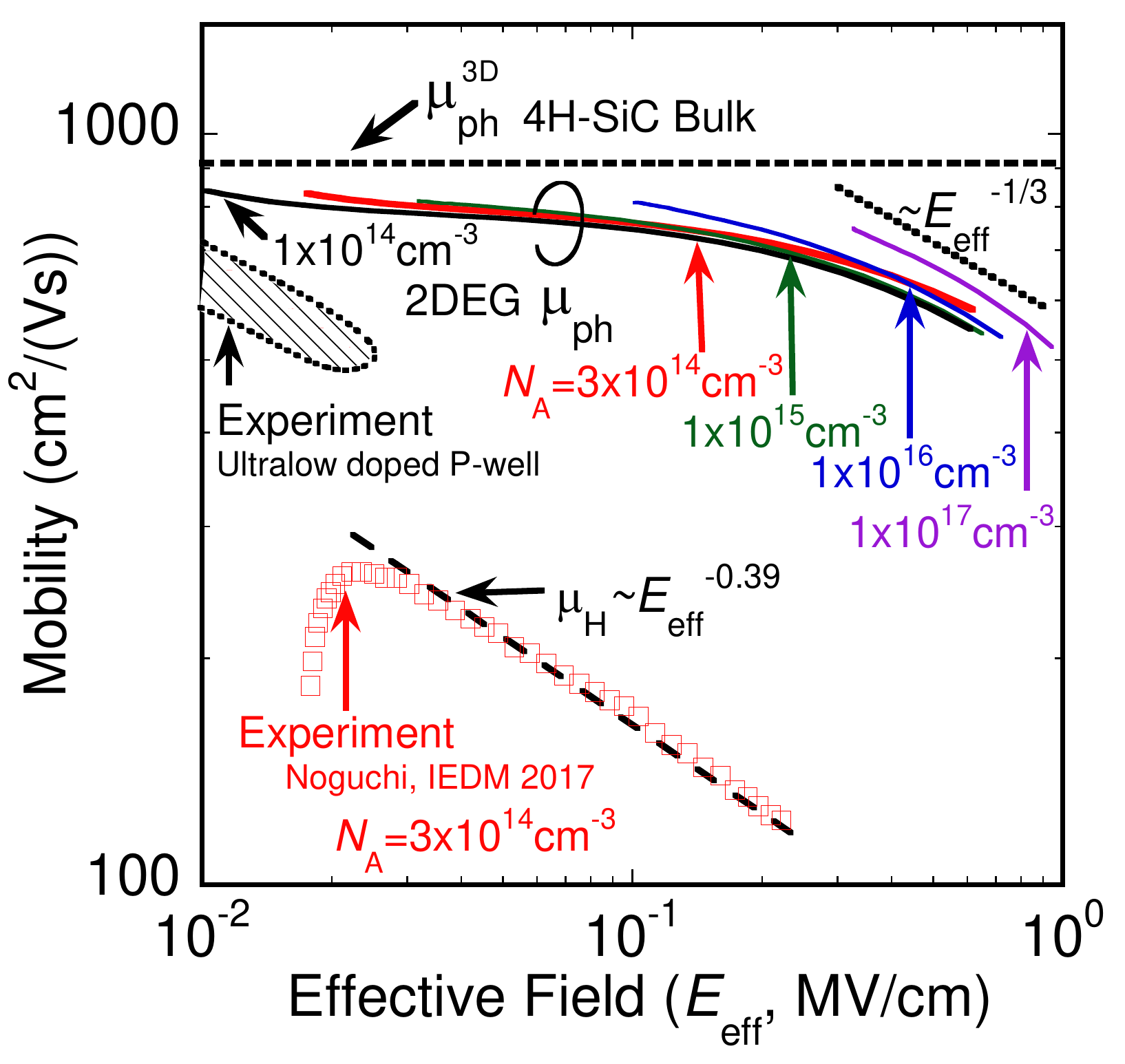}
\caption{A comparison between the calculated phonon-scattering-limited mobility ($\muph$)
and the experimentally obtained free electron mobility ($\muF$) for MOSFETs on a low-doped ($\Na=3\times 10^{14}$ cm$^{-3}$) P-well
and a ultralow-doped ($\Na < 1 \times 10^{14}$) 
P-well, where $\Na$ denotes the acceptor density of the P-well \cite{Noguchi_2017, Sometani_2019}.
The dependence of the $\muF$ on the $\Eeff$ reported by Sometani {\it et al.} is estimated by assuming that the $\Na$ of 
the ultralow-doped P-well is  $10^{13}$ cm$^{-3}$. }
\label{fig:mobph}
\end{figure}
First, the $\muph$ in SiC MOSFETs was calculated to examine how it compares with the experimentally 
observed mobility \cite{Noguchi_2017,Sometani_2019}.
In this calculation, we account for two important phonon scattering mechanisms:
 acoustic phonon scattering and inter-valley nonpolar optical phonon scattering \cite{Masaki_1989,Iwata_2001}.
The parameters for phonon scattering potentials are taken from previous work
 that investigated the bulk mobility of 4H-SiC \cite{Iwata_2001}.
Fig.~\ref{fig:mobph} shows the comparison of experimental results and the calculated 
 $\muph$'s for SiC MOSFETs on P-wells, which have various $\Na$'s \cite{Noguchi_2017, Sometani_2019}.
In Fig.~\ref{fig:mobph}, the horizontal axis shows $\Eeff$ defined in Eq.~\ref{eq:Eeff}.
We note that the approximated analytical solution for the $z$-directional 
envelope function ($\xi(z)$) of the 2DEG can be scaled by the value of $\Eeff$ 
taking $\eta=11/32$ in Eq.~\ref{eq:Eeff} \cite{Stern_1967}.
As a result, in the quantum limit, the $\muph$ in MOSFETs is expected to be proportional to $\Eeff^{-1/3}$ \cite{Takagi_1994_1,Takagi_1994_2}. 
In fact, the ``universal mobility'' (the dominant component of the mobility in Si MOSFETs
scaled by the value of $\Eeff$) of electrons in Si MOSFETs is set by the $\muph$ 
for the $\Eeff$ range from 0.05 to 0.5 MV/cm \cite{Takagi_1994_1}.
In this range of $\Eeff$, the mobility in Si MOSFETs is approximately proportional to $\Eeff^{-0.3}$.
Fig.~\ref{fig:mobph} shows that the calculated values of $\muph$ show a universal behavior: the
calculated $\muph$ shows the same dependence on $\Eeff$, irrespective of the values of $\Na$.
However,the expected power-law behavior of $\muph$ against the $\Eeff$ 
 $(\muph \sim \Eeff^{-1/3})$ is only observed in the relatively high values of $\Eeff$  ($\Eeff >$ 0.1 MV/cm).
For smaller values of the $\Eeff$, the calculated $\muph$ approaches the bulk mobility of 4H-SiC asymptotically 
as the $\Eeff$ decreases; this occurs along with a transition of the electronic structure from
a quasi-2DEG to a three-dimensional electron gas (3DEG), which is localized around the interface.
These calculated results are consistent with the experimental results that 
showed the highest measured mobility of a SiC MOSFET on an ultralow-doped P-well, as shown 
in Fig.~\ref{fig:mobph} \cite{Sometani_2019}.
As the $\Eeff$ increases toward 1 MV/cm, the calculated value of the $\muph$ 
decreases down to around 50 \% of the bulk mobility;
this decrease is due to the shrinkage of the width of $\xi(z)$ or the thickness of the 2DEG.
By contrast, the experimental values of the $\muF$ 
for a low-doped P-well (with $\Na=3\times 10^{14}$ cm$^{-3}$) 
is much smaller than the calculated values of the $\muph$.
As described previously,in the case of low-doped P-wells,
the experimental $\muF$ is approximately proportional to $\Eeff^{-0.39}$ ($\muF \sim \Eeff^{-0.39}$);
the exponent in this expression is close to the theoretically obtained exponent of $-1/3$ 
under the assumption of the quantum limit \cite{Price_1981}.
However, for small values of $\Eeff$, the exponent $\alpha$ in the relationship between
the calculated $\muph$ and the $\Eeff$, $\muph \sim \Eeff^{\alpha}$, 
is not found to be $-1/3$; the diminish of the effect of $\Eeff$ on $\muph$ for small values of $\Eeff$,
is caused by the transition of the electronic structure from a 2DEG to a 3DEG.
In brief, as is clearly shown in Fig.~{fig:mobph}, the experimental relationship $\muF \sim \Eeff^{-0.39}$
could not reproduced by the calculation of the $\muph$.\par

We also calculated the temperature dependence of the mobility;
as expected, the temperature dependence of the $\muph$ of the 2DEG is almost equivalent 
to that of the bulk mobility. The $\muph$ shows a rapid increase as the temperature decreases.
Experimentally obtained observations of temperature dependence of the mobility at low temperatures are limited 
\cite{Noguchi_2019,Amano_2017,Hirai_2019}.\par

In view of above findings, we conclude that the mobility in SiC MOSFETs on low-doped P-wells is
limited by a scattering mechanism other than phonon scattering.\par
%
%
\begin{figure}[hbtp]
\centering
\includegraphics[width=8.6cm]{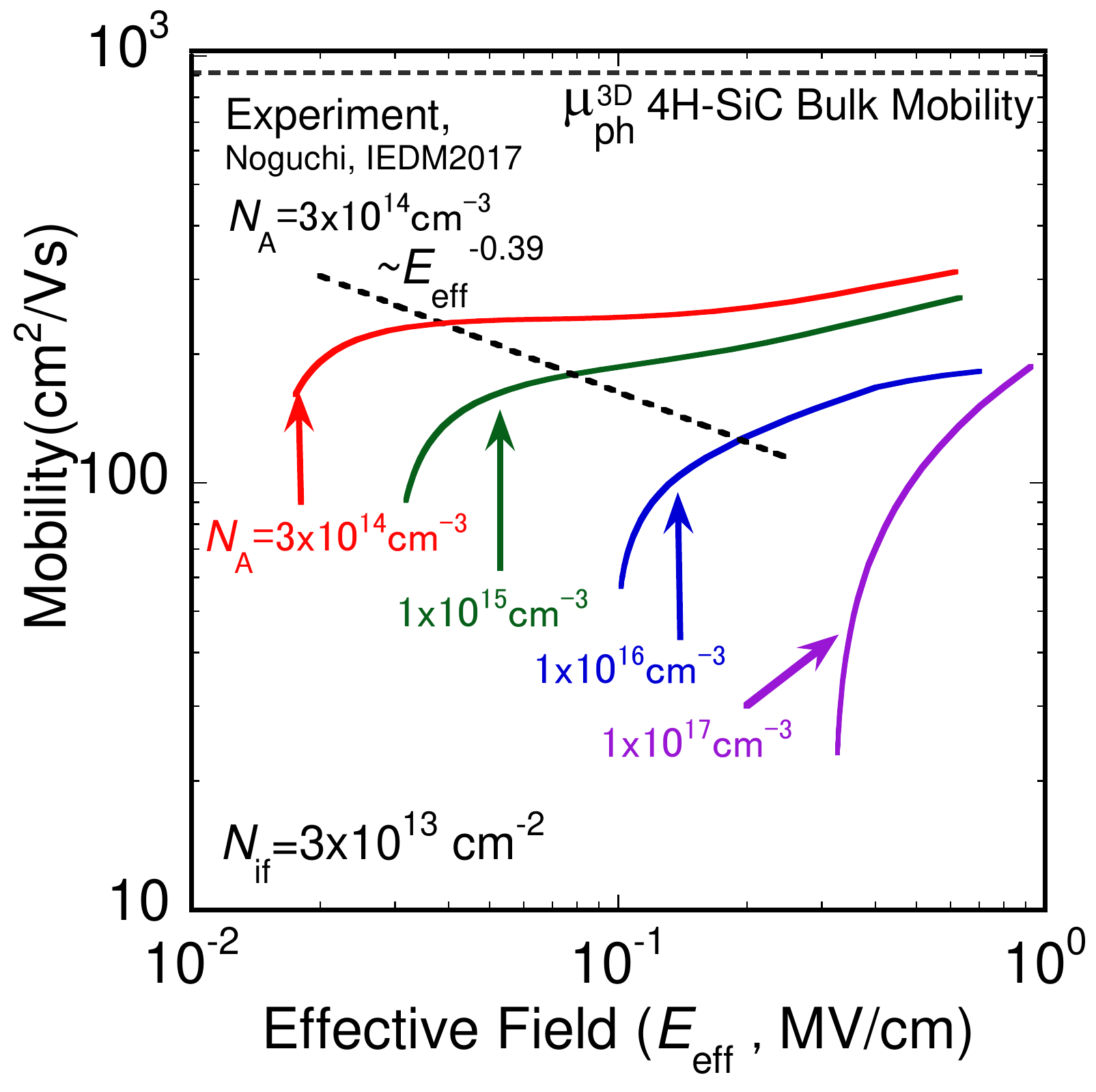}
\caption{A comparison of the calculated values of  Coulomb-scattering-limited mobility,
$\muC$, in SiC MOSFETs 
on P-wells with various values of $\Na$ with the experimentally
obtained free electron mobility, $\muF$, (in  the experimental work, $\Na=3\times 10^{14}$ cm$^{-3}$)
\cite{Noguchi_2017}.
}
\label{fig:Coulomb} 
\end{figure}
In order to identify the dominant scattering mechanism,
we examine whether Coulomb-scattering-limited mobility, $\muC$, can reproduce the experimental mobility 
observed in SiC MOSFETs on a low-doped P-well.
As a preliminary investigation, we evaluate the effect of ionized impurities in the depletion region
on the inversion layer mobility of SiC MOSFETs using simple analytical model. 
The calculated results indicate that the effect of ionized impurities in the depletion region can be neglected in this study.
Therefore, to simplify the problem, we assumed that there exist a high density of  Coulomb scattering centers only at the interface region. 
We examine a sufficient number of subbands in the calculation of 
the $\muC$ to accurately describe the change in the electronic structure from three-dimensional to quasi-two-dimensional behavior.
In addition, we have integrated the screening theory for the multi-subband calculation of the $\muC$ 
\cite{Price_1981,Price_1982, Hatakeyama_icscrm_19,Tanaka_2020}.\par
Here, we briefly outline the Coulomb scattering theory of a 2DEG in order to
better understand the obtained results of the $\muC$.
The Fourier-transformed unscreened Coulomb potential $\ptC(k,z)$
is given by:
\begin{equation}
\ptC(k,z)=
\frac{e}{\epss}\frac{\re^{-k z}}{2k},
\label{eq:Coulomb}
\end{equation}
where an elementary charge is assumed to be located at the origin of the cylindrical
coordinate system.
From Eq.~\ref{eq:Coulomb}, it can be seen that $\ptC(k,z)$ diverges, as $k$ tends to zero.
This behavior of $\ptC(k,z)$ is referred to as  an infrared divergence, which is 
due to the Coulomb interaction acting over long distances.
In the calculation of the mobility, the infrared divergence corresponds to 
the scattering frequency diverging to the positive infinity
as the scattering angle approaches to zero.
The Coulomb scattering is thus dominated by small-angle scattering.
Since the Coulomb scattering center is surrounded by free electrons in
the MOS channel, the Coulomb potential is screened.
The screened Coulomb potential $\ptscC(k,z)$ 
is expressed by the sum of the unscreened potential $\ptC(k,z)$ and 
the potential arises from the induced charges in the 2DEG.
In the multi-subband screening theory, a screening constant for the $i$-th suband, $s_i$,
is defined:
\begin{equation}
s_i=\frac{e^2}{2 \epss}
\frac{mn_v}{\pi \hbar^2}
\left[1-
\exp \left[-\frac{n_i \pi \hbar^2}{mn_v \kb T}\right]
\right],
\label{eq:si}
\end{equation}
where $n_i$, $m$, and $n_v$ are the electron density of the $i$-th subband,
the effective mass of an electron, and the number of valleys, respectively.\par
At the quantum limit at room temperature, (or, more precisely, at the non-degenerate limit \cite{Daviesbook}),
Eq.~\ref{eq:si} can be simplified; the screening constant of the lowest subband
$s_1$ is found to be proportional to the $\nfree$:
\begin{equation}
s_1=\frac{e^2}{2 \epss}
\frac{\nfree}{\kb T},
\label{eq:RTs1}
\end{equation}
where $n_1$ is replaced with $\nfree$ because
all the free electrons in the MOS channel
occupy the lowest subband in quantum limit.
We can also derive the analytical expression for
$\ptscC(k,z)$ in the quantum limit,
as follows:
\begin{equation}
\ptscC(k,z)=
\frac{e}{\epss}\frac{\re^{-k z}}{2(k+F_{11}(k)s_1)},
\label{eq:scC}
\end{equation}
where $F_{11}$ is a form factor of the lowest subband.
In general, a form factor $ F_{ij}(k)$ is defined as:
\begin{equation}
F_{ij}(k)
=\iint |\xi_i(z_1)|^2 |\xi_j(z_2)|^2 e^{-k|z_2-z_1|}dz_1 dz_2,
\label{eq:Fij}
\end{equation}
where $\xi_i(z)$  is  the envelope function 
of the $i$-th subband.
The function $\ptscC(k,z)$, as expressed in Eq.~\ref{eq:scC}, 
rapidly increases  as $k$ tends to zero. 
Accordingly, the scattering due to $\ptscC(k,z)$
is also dominated by the contribution close to $k=0$, which corresponds
to small-angle scattering.
The frequency of the small-angle scattering is subject to
a screening effect, because the limit value at zero ($\ptscC(0,z)$) is
proportional to the inverse of the screening constant ($\sim 1/s_1$) or
the free carrier density ($\sim 1/\nfree$), as shown in  Eq.~\ref{eq:RTs1}.\par
By applying multi-subband scattering theory to the $\ptscC(k,z)$,
the values of $\muC$ for MOSFETs on P-wells with different values of $\Na$ were calculated.
The calculated values of $\muC$ are shown in Fig.~\ref{fig:Coulomb}.
In the calculation, the area density of Coulomb scattering centers
at the interface, $N_\mathrm{if}$, was treated as a fitting parameter.
To reproduce the measured values of $\muF$, which were as high as 200 cm$^2$/Vs,
$N_\mathrm{if}$ was taken to be $3 \times 10^{13}$ cm$^{-2}$.
The $\Eeff$ was calculated according to Eq.~\ref{eq:Eeff}, where
$\Ns$ was varied between $1 \times 10^{10}$ cm$^{-2}$ and  $1 \times 10^{13}$ cm$^{-2}$.
As is shown in Fig.~\ref{fig:Coulomb}, the value of $\muC$ increases as the
$\Eeff$ increases, irrespective of the value of $N_\mathrm{A}$ of the P-well. 
This sharp increase in the calculated $\muC$ for the smaller values of $\Eeff$  in each 
of $\Eeff$-$\muC$ curves 
is due to the fact that Coulomb scattering is dominated by small-angel scattering, 
the frequency of which can be reduced by the increase of 
screening effect, which is in turn caused by an increase in $\nfree$.
The increase in the $\Eeff$ also enhances the frequency of
large-angle scattering as it leads to a reduction of
the average distance between electrons and scattering centers:
The frequency of large angle scattering increases
due to an increase in the exponential term in the expression for $\ptscC(k,z)$ in Eq.~\ref{eq:scC},
which is accompanied by a decrease in the average distance. 
However, it is seen that the screening effect dominates the relationship between
 $\Eeff$ and $\muC$: {\it i.e.} $\muC$ increases at  $\Eeff$ increases.
The dashed line in Fig.~\ref{fig:Coulomb} indicates  the experimentally measured values of
 $\muF$ for a MOSFET on low-doped P-well  (for which $N_\mathrm{A}=3\times10^{14}$ cm$^{-3}$).
It is found that the measured values of $\muF$ decreases as the values of $\Eeff$ increases 
when $\nfree$ is larger than  $\Ndpl$.
Thus, considering the observed differences in the dependence of the experimental and the calculated
results on the  $\Eeff$,
we can conclude that the dominant contribution to the measured values of the $\muF$
 do not originate from the Coulomb scattering.\par
Furthermore we should also note that the assumed $N_\mathrm{if}$ in the mobility calculation
is not consistent with the measured value of  threshold voltage ($\Vth$) for the SiC MOSFETs. 
In the experiments, the measured value of $\Vth$ for a SiC MOSFET is typically
less than 5V \cite{Hatakeyama_2017, Hatakeyama_2019}.
These experimental results show that the area density of the
fixed charges at the interface is negligibly small compared with the assumed value of $N_\mathrm{if}$
used in the mobility calculation.
In order for this situation to physically occur, the electric charges at interface would have to
approximately cancel out: around half of the scattering centers would have to be negatively charged,
and others would have to be positively charged. This condition is not physically improbable.\par
Before considering the possible candidates for the dominant scattering mechanism,
we re-examine why  $\muC$ is a monotonic increasing function of $\Eeff$.
We note that according to Eq.~\ref{eq:Eeff},
an increase in $\Eeff$ is caused by an increase in $\nfree$ in a MOSFET. 
The value of $\muC$ increases as the  value of $\nfree$ increases 
due to the screening effect (because the small-angle scattering dominates Coulomb scattering).
This is due to the fact that the Coulomb interaction is an interaction that occurs over a long (or, infinite) distance.
Thus, we can conclude that the potential of the dominant scattering mechanism must be short-ranged in nature:
it must decay faster than  $\sim 1/\rtd$, where $\rtd$  is a magnitude of the three-dimensional position vector.\par
Here, in order to establish the cause of the low mobility, we consider other scattering mechanisms. 
We can exclude the phenomenon of surface-roughness scattering as the cause of the decreased mobility. Surface-roughness-scattering-limited mobility is proportional to the square of the inverse of the effective electric field.
Thus, in the relevant effective electric field range for a MOSFET on a low-doped P-well,
 surface-roughness-scattering-limited mobility is too small to reproduce the experimentally observed values of $\muF$.\par
It is assumed that the potential of the dominant scattering center that causes the decrease in mobility 
acts only over a short distance (a short-range scattering potential). 
Furthermore, the following two assumptions can be made regarding the scattering centers of the dominant mechanism: 
(1) The scattering centers are neutral.
(2) The scattering centers are predominantly located at the interface of the materials. 
The first assumption is based on the measured $\Vth$ of the SiC MOSFETs. 
The second assumption is implied by the stability of defects, which are the origin of the scattering centers.
At the MOS interface, there exists a lattice mismatch. 
Therefore, at the interface, there exists residual strain and it is probable that dangling bonds are present. 
This may lead to a high density of defects at the interface. 
On the bulk side of the interface, the density of defects is likely to be limited.\par
When a defect is formed in a lattice structure, the distribution of charge density deviates from 
the distribution of  that would be expected in the case of a perfect lattice.
The perturbation potential of a defect, $\delta V(\brc)$, can be
expressed by the deviation of the charge density, $\delta \rho(\brc)$, according to,
\begin{equation}
\delta \phi(\brc)=\frac{1}{4 \pi \varepsilon_{0}}
 \int \frac{\delta \rho\left(\brc^{\prime}\right)}{\left|\brc-\brc^{\prime}\right|} d V^{\prime},
\end{equation}
where $\brc$ is the position vector in the Cartesian coordinate system.
We apply the multipole expansion technique to this defect potential, as follows:
\begin{equation}
\delta \phi(\brc)=
\frac{1}{4 \pi \varepsilon_{0} \rc} 
\sum_{\ell=0}^{\infty} \frac{1}{\rc^{\ell}} 
\int \delta\rho\left(\brc^{\prime}\right) 
P_{\ell}\left(\cos \theta^{\prime}\right) \rc^{\prime \ell} d V^{\prime} \equiv \sum_{\ell=0}^{\infty} \delta \phi_{\ell}(\brc),
\end{equation}
where $P_\ell(x)$ is the $\ell$-th Legendre polynomial. 
The first isotropic term of the multipolar expansion of $\delta \phi_0(\brc)$ is 
the Coulomb potential for the sum of the charges in the defect area. 
We neglect this first term because the defect is assumed to be neutral (as stated above).
Thus, the second term becomes the dominant term in the defect potential.
The second term, $\delta \phi_1(\brc)$, is defined as follows:
\begin{equation}
\delta \phi_1(\brc)=\frac{1}{4 \pi \varepsilon_{0} \rc^3}\brc \cdot \bpdp,
\end{equation}
where  $\bpdp$ is the electric dipole moment of a defect defined as,
\begin{equation}
\bpdp=\int \brc \delta \rho(\brc) dV.
\end{equation}
To calculate the value of  $\muF$ for  SiC MOSFETs, we consider a cylindrical coordinate system.
Furthermore, the dielectric constant is changed from $\varepsilon_0$ to  $\epss$,
as it is the interaction between the dipole and free electrons in semiconductor 
that we must consider here.
The potential of the electric dipole located at the MOS interface in cylindrical coordinate system is 
then expressed as,
\begin{equation}
\phi_{\mathrm{dp}}(r,z)=\frac{1}{4\pi\epss}
\frac{z}{(r^2+z^2)^{3/2}}p_{\mathrm{dp}},
\label{eq:dipole}
\end{equation}
where $p_{\mathrm{dp}}$ is the norm of the dipole moment $\bm{p}_{\mathrm{dp}}$.
For simplicity, $\bm{p}_{\mathrm{dp}}$ is assumed to be orthogonal to the MOS interface.
Eq.~\ref{eq:dipole} shows that the potential of a dipole acts over a short distance:
the potential is seen to be inversely proportional to the square of the distance ($\sim 1/\rtd^2$) ,
whereas a Coulomb potential is inversely proportional to the distance ($\sim 1/\rtd$).\par

Next, we examine the effect of short-ranged dipole potential on electron scattering.
The Fourier transformed potential of a dipole, $\phi_{\mathrm{dp}}(k,z)$, is 
given by,
\begin{equation}
\phi_{\mathrm{dp}}(k,z)=\frac{p_{\mathrm{dp}}}{2\epss}
\re^{-k z}.
\label{eq:kdipole}
\end{equation}
From Eq.~\ref{eq:kdipole}, we can see that $\phi_{\mathrm{dp}}(k,z)$ converges to a constant
as $k$ tends to zero, whereas  $\ptC (k,z)$ diverges (see Eq.~\ref{eq:Coulomb}).
Since the dipole scattering center is surrounded by free electrons in
the MOS channel, the dipole potential is screened.
The screened dipole potential, $\phi_{\mathrm{scdp}}(k,z)$, 
can be calculated using multi-subband screening theory in the same way as the screened Coulomb
potential.
At the quantum limit at room temperature, the analytical expression of the screened dipole potential, $\phi_{\mathrm{scdp}}(k,z)$,
can be expressed as,
\begin{equation}
\phi_{\mathrm{scdp}}(k,z)=
\frac{p_{\mathrm{dp}}}{2\epss}\frac{k\re^{-k z}}{(k+F_{11}s_1)}.
\label{eq:scdp}
\end{equation}
From Eq.~\ref{eq:scdp}, we can see that $\phi_{\mathrm{scdp}}(k,z)$ tends to zero 
for small values of $k$.
The reduction in the amplitude of the potential around $k=0$ of the dipole potential
induces a  reduction in the statistical frequency of small-angle scattering.
As a result, the ratio of small-angle scattering to the total scattering is
reduced compared with ratio obtained in the case of Coulomb scattering.
As discussed above, the increase in $\Eeff$ induced by
the increase in $\nfree$ simultaneously causes a  suppression in the frequency of the small-angle
scattering and an increase in the frequency of large-angle scattering.
Thus, in the case of dipole scattering,
an increase in $\Eeff$ tends to increase the frequency of the total scattering.
To examine the dependence of  $\mudp$ on  $\Eeff$ and how this dependence
is affected by these different types of scattering, numerical analysis is indispensable.\par
\begin{figure}[hbtp]
\centering
\includegraphics[width=8.6cm]{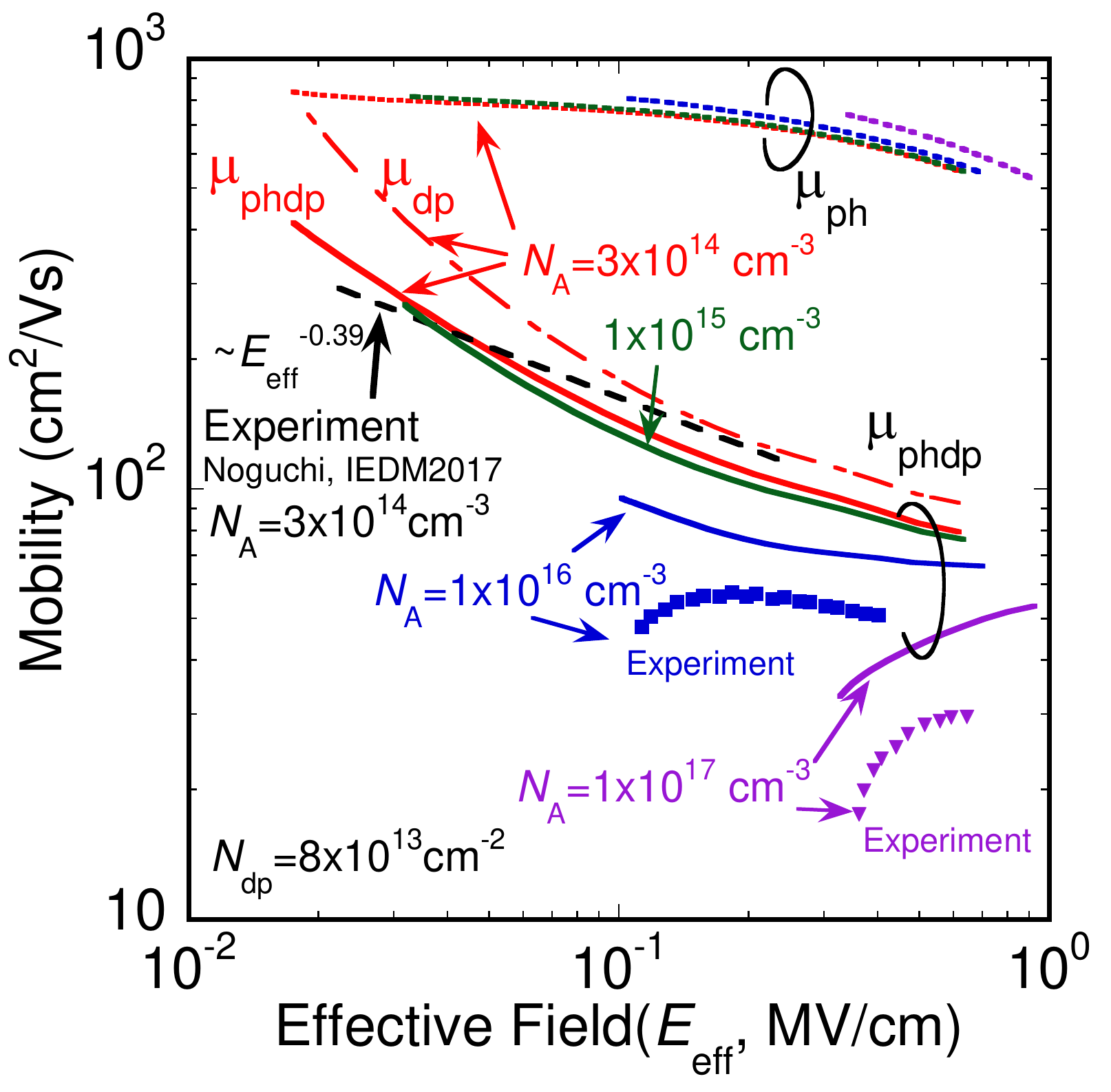}
\caption{The calculated $\Eeff$ dependence of 
the calculated dipole scattering-limited mobility combined
with phonon scattering-limited mobility ($\mudpph$), 
the dipole scattering-limited mobility ($\mudp$),
phonon scattering-limited mobility ($\muph$),
and the measured  $\muF$ of MOSFETs on a low-doped P-well ($\Na=3\times 10^{14}$ cm$^{-3}$).
The $\mudpph$ and the measured values of $\muF$ on a medium-doped P-well  ($N_\mathrm{A}=1\times10^{16}$ cm$^{-3}$)
and on a high-doped P-well ($N_\mathrm{A}=1\times10^{17}$ cm$^{-3}$) are also shown \cite{Noguchi_2017}.
}
\label{fig:dipole}
\end{figure}

By using the screened dipole potential, $\phi_{\mathrm{scdp}}(k,z)$, we
calculated the dipole-scattering-limited mobility, $\mudp$, according 
to the multi-subband scattering theory of 2DEGs.
In this calculation, the magnitude of the dipole moment, $\pdp$,
and the area density of the dipole, $\Ndp$, 
are treated as fitting parameters. 
The $\pdp$ is assumed to be $e\cdot d$, 
where $d$ is the distance between the two point charges, which is assumed to be 1 nm.
The value of $\Ndp$ that best reproduces the experimental results is found to be  
 $8\times10^{13}$ cm$^{-2}$.
In Fig.~{\ref{fig:dipole}}, we show the calculated dipole-scattering-limited mobility combined
with phonon-scattering-limited mobility, $\mudpph$, the dipole-scattering-limited mobility, $\mudp$,
and phonon-scattering-limited mobility, $\muph$, on a low-doped P-well ($\Na=3\times 10^{14}$ cm$^{-3}$).
The experimentally obtained $\muF$-$\Eeff$ relation is also shown in Fig.~{\ref{fig:dipole}}. 
The values of $\mudpph$ and the experimental results on a medium-doped P-well  ($N_\mathrm{A}=1\times10^{16}$ cm$^{-3}$)
and on a high-doped P-well ($N_\mathrm{A}=1\times10^{17}$ cm$^{-3}$) are also shown in Fig.~{\ref{fig:dipole}}.\par

In the case  of the  low-doped P-well, the calculated value of $\mudpph$ decreases as the value of $\Eeff$ increases.
The calculated values of $\mudpph$ on a low-doped P-well match the experimental results well.
The calculated values of $\mudpph$ on a medium-doped P-well and a high-doped P-well 
do not match the experimental results, at least quantitatively.
However, the gradients of both the calculated $\mudpph$-$\Eeff$ curves and 
the experimentally obtained $\muF$-$\Eeff$ curves similarly change from negative to 
positive as the the doping density of the P-well increases.
This variation in the gradient of the $\mudpph$-$\Eeff$ curve with increasing doping density 
is a distinctive feature of the experimentally obtained $\muF$ in SiC MOSFETs. 
The calculated $\mudpph$ successfully reproduced this feature, which 
can be understood by considering the change in the magnitude of the analytical 
dipole scattering potential given in Eq.~\ref{eq:scdp}.
As the $\nfree$ increases, the the values of $\Eeff$ and $s_1$ increase according to
the Eqs.~\ref{eq:Eeff} and \ref{eq:RTs1}, respectively.
As the doping density of the P-well increases, the effect of $\nfree$
on $\Eeff$ decreases according to Eq.~\ref{eq:Eeff}, whereas that on the screening effect is unchanged.
Accordingly, as  the doping density of the P-well increases, the screening effect dominates the 
$\mudpph$-$\Eeff$ relation. As a result, as the $\Eeff$ increases, the mobility increases in the case of 
the high-doped P-well.
\par
\begin{figure}[hbtp]
\centering
\includegraphics[width=8.6cm]{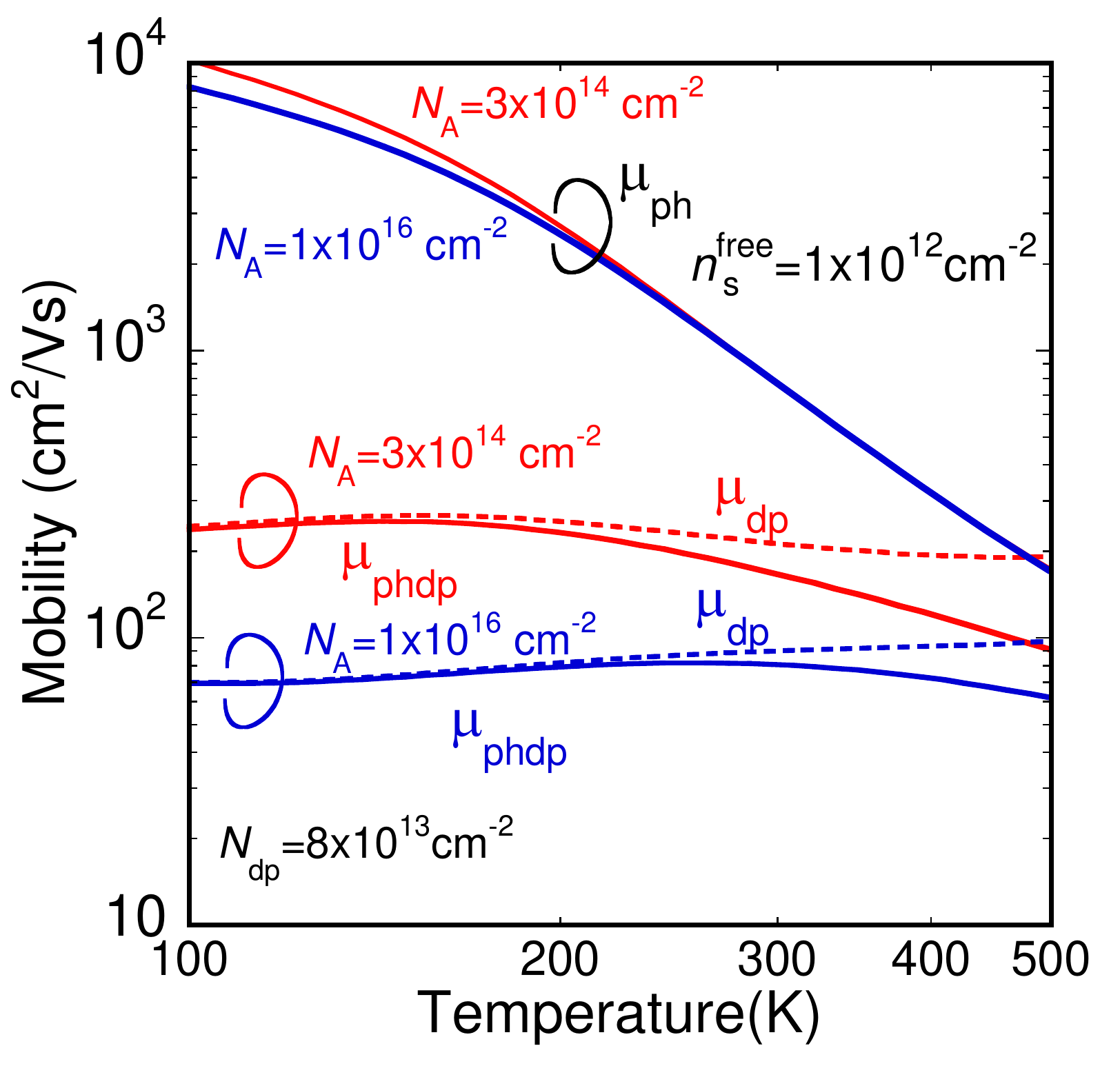}
\caption{The calculated temperature dependence of $\muphdp$, $\mudp$ 
and $\muph$ in the case of a low-doped P-well and a medium-doped P-well }
\label{fig:dptemp}
\end{figure}
In Fig.~\ref{fig:dptemp},
we show the calculated temperature dependence of  $\mudpph$, $\muph$ and
$\mudp$ for the SiC MOSFETs on a low-doped P-well and a medium-doped P-well.
The temperature dependence of $\mudpph$ in the case of a low-doped P-well and a medium-doped P-well
is limited.
The observed temperature dependence of $\mudpph$ qualitatively agrees with previous experimental 
results \cite{Noguchi_2019,Amano_2017,Hirai_2019}.
In the case of the low-doped P-well at temperatures above room temperate, 
the value of $\mudpph$ decreases as the temperature increases.
This temperature dependence is caused by the rapid decreases in the value of $\muph$, as shown in Fig.~\ref{fig:dptemp}.
The temperature dependence of $\mudp$ is limited regardless the doping density of the P-well.
This small temperature dependence of  $\mudp$ is caused by the cancellation of two effects on the screened dipole
potential, $\phi_{\mathrm{scdp}}(k,z)$: 
the first being an increase in the magnitude of the potential accompanied by the decrease of the average energy 
with decreasing temperature,
and the second effect being a decrease in the magnitude of the potential caused by
the increase of the screening  effect that occurs with a
decrease in the temperature according to Eqs.~\ref{eq:scdp} and \ref{eq:RTs1}.
\par
As described above, the calculated values of $\mudp$ closely match the experimental 
observations in terms of their dependence on $\Eeff$, $\Na$ and temperature.
This indicates that the scattering potential of the defects at the SiC MOS interface 
acts over a short distance (a short-range potential) and that there exist a 
high density of dipoles at the interface.
Thus, it can be concluded that a dominant scattering mechanism affecting the mobility of SiC MOSFETs
is due to dipole scattering.
Next, we consider further evidence that may help to verify this hypothesis.\par
We first discuss the relation among the calculated $\mudp$, the assumed value of $\pdp$, and the fitted value of $\Ndp$.
According to the scattering theory of the mobility calculation,
the calculated value of the $\mudp$ is inversely proportional to the value of
 $\pdp^2 \Ndp$, because the scattering probability is proportional to the product of the $\Ndp$ and
the square of the dipole potential.
Therefore, for a constant value of $\pdp^2 \Ndp$, $\Ndp$ can be changed 
to a given value of $\pdp$ without necessarily altering the value of $\mudp$.
By comparing the calculated mobility with observations from experimental work,
it is only possible to gain information regarding the value of $\pdp^2 \Ndp$.\par
Here, we consider the origin of the dipoles at the interface.
If the dipoles at the interface are directly caused by the defects by themselves,
the value of $\pdp$ of the defect may be smaller than the assumed value of $\pdp$ 
(which was taken to be equal to $e\cdot d$ with $d=1$ nm).
Thus, if we replace the assumed $\pdp$ with a smaller value, 
the value of $\Ndp$ should increase in order to maintain the constant value of $\pdp^2 \Ndp$
(and therefore the agreement with the experimental work).
However, there exists an upper limit for the value of $\Ndp$;
it cannot be greater than the area density of the surface atoms of 4H-SiC, which is around $2.1 \times 10^{15}$ cm$^{-2}$.
If the actual value of $\pdp$ is less than 1/3 of the $\pdp$ assumed here,
the required value of $\Ndp$ would be comparable the area density of the surface atoms.
Furthermore, such a value would be inconsistent with the numerous results obtained 
from transmission electron microscopy (TEM) observations of SiC/SiO$_2$ interfaces,
where the arrangement of atoms on the SiC side of the interface are retained in the vicinity of 
the interface \cite{Peizhi2014,taillon_2015,hosoi2011,hatakeyama2011}.
The formation energy of defects that are large in size, such as carbon clusters,
limits their area density.
Indeed, it has been established by TEM analysis and electron spin resonance measurements
that the density of carbon clusters at SiC/SiO$_2$ interfaces 
is smaller than $2 \times 10^{13}$ cm$^{-2}$ \cite{kagoyama2019,umeda2018}.
Thus, the hypothesis that the dipole moment is directly due to the presence of
the defects is not consistent with the observed interface structure.\par
Here, we focus on the electronic structure and the charge distribution unique
to the interface region. At the MOS interface, translational symmetry
is broken. A charge transfer layer (or dipole layer) is formed at the interface 
via the mutual penetration of the wave function from both sides of the interface \cite{tersoff1984}.
This charge transfer layer affects the band alignment between the two material 
through dipoles formed by charge transfer at the interface. 
This phenomenon is sensitive to the local structure at the interface.
Indeed, it has been reported that the shift in the band alignment of
SiC and oxide can be induced by performing hydrogen annealing after the gate oxidation \cite{watanabe2011,hosoi2012}.
Based on these observations, we propose the hypothesis that the small defects at the interface
cause a  rearrangement of the charge transfer layer, which leads to the formation of a large dipole.
To verify this hypothesis, further experimental and theoretical investigations into the 
MOS interface must be undertaken, including first-principles calculation of
the electronics structures at the interface.\par
In summary, the cause of the low free electron mobility in SiC MOSFETs was investigated
considering the scattering theory of 2DEGs.
By comparing the calculated results and  experimental results, we have
shown that neither phonon scattering nor Coulomb scattering can be the cause of 
the low mobility.
By considering the threshold voltages and the nature of the effective field dependence of 
the mobility in SiC MOSFETs,
it is established that the scattering centers of the dominant mechanism are electrically neutral and 
exhibit a short-range scattering potential.
Based on the charge distribution around a neutral defect
at the interface, we conclude that the scattering center is
most likely to be an electric dipole.
The calculated dipole-scattering-limited mobility shows 
similar effective field dependence to that observed in experimental work.
Furthermore, the origin of the dipoles at the interface is discussed based on the 
extracted parameters from the scattering theory and physics of the electronic
strictures of the interface.
We concluded that the scattering that occurs  due to a high density of electric dipoles at the interface 
is the most likely cause of the low mobility of SiC MOSFETs.

\begin{acknowledgments}
This work was supported by JSPS KAKENHI Grant Number JP19K04494 and
This work was supported by MEXT-Program for Creation of Innovative Core Technology for Power Electronics Grant Number JPJ009777.
\end{acknowledgments}
\nocite{*}
\bibliography{dipole_bib}

\end{document}